\documentclass[preprint3]{aastex}

\usepackage{fixltx2e}
\usepackage{ amsmath } 
\usepackage{ amssymb }

\newcommand{\beq}{\begin{equation}}
\newcommand{\calR} {{\cal R}}
\newcommand{\eeq}{\end{equation}}

\newcommand{\LJ} {L_{\rm J}}
\newcommand{\msun}{M_\odot}
\newcommand{\pcc} {{\rm ~cm}^{-3}}
\newcommand{\psc} {{\rm ~cm}^{-2}}
\newcommand{\rholin} {\rho_{\rm 1D}}

\newcommand{\rhosh} {\rho_{\rm 2D}}
\newcommand{\rhosph} {\rho_{\rm 3D}}
\newcommand{\tff} {\tau_{\rm ff}}
\newcommand{\tffbh} {\tau_{\rm ff,BH}}
\newcommand{\tffsh} {\tau_{\rm ff,2D}}
\newcommand{\tffsph} {\tau_{\rm ff,3D}}
\newcommand{\tffone} {\tau_{\rm ff,1}}
\newcommand{\tfftwo} {\tau_{\rm ff,2}}

\shorttitle{Free-fall time of sheets and filaments}
\shortauthors{Toal\'a, V\'azquez-Semadeni \& G\'omez}
\begin{document}
\title{The free-fall time of finite sheets and filaments}
\author{Jes\'us A. Toal\'a, Enrique V\'azquez-Semadeni \& Gilberto C. G\'omez}
\affil{Centro de Radioastronom\'\i a y Astrof\'\i sica, Universidad Nacional
Aut\'onoma de M\'exico, Campus Morelia} 
\affil{Apartado Postal 3-72, 58090, Morelia, Michoac\'an, M\'exico}
\affil{ }
\affil{To appear in The Astrophysical Journal}

\begin{abstract}
Molecular clouds often exhibit filamentary or sheet-like shapes. We
compute the free-fall time ($\tff$) for finite, uniform,
self-gravitating circular sheets and filamentary clouds of small but
finite thickness, so that their volume density $\rho$ can still be
defined. We find that, for thin sheets, the free-fall time is larger
than that of a uniform sphere with the same volume density by a factor
proportional to $\sqrt{A}$, where the aspect ratio $A$ is given by
$A=R/h$, $R$ being the sheet's radius and $h$ is its thickness.  For
filamentary clouds, the aspect ratio is defined as $A=L/\calR$, where
$L$ is the filament's half length and $\calR$ is its (small) radius, and
the modification factor is a more complicated, although in the limit of
large $A$ it again reduces to nearly $\sqrt{A}$. We propose that our result
for filamentary shapes naturally explains the ubiquitous configuration
of clumps fed by filaments observed in the densest structures of
molecular clouds. Also, the longer free-fall times for non-spherical
geometries in general may contribute towards partially alleviating the
``star-formation conundrum'', namely, that the star formation rate in
the Galaxy appears to be proceeding in a timescale much larger than the
total molecular mass in the Galaxy divided by its typical free-fall
time. If molecular clouds are in general formed by thin sheets and long
filaments, then their relevant free-fall time may have been
systematically underestimated, possibly by factors of up to one order of
magnitude.

\end{abstract}


\section{Introduction} \label{sec:intro}

The so-called free-fall time is one of the most important quantities in
astrophysics. For a spherical object of mass $M$ and radius $R$, this
timescale is given by \citep[see, e.g.,][]{BT87}
\begin{equation}
\tff \equiv \sqrt{\frac{\pi^2 R^3}{8GM}} = \sqrt{\frac{3 \pi}{32
G \rho}}; 
\label{eq:tff}
\end{equation}
where in the second equality we have introduced the volume density
defined by
\beq
\rho(M,R) = \frac{3M}{4 \pi R^3}.
\label{eq:ro3D}
\eeq

The timescale $\tff$
has the interesting property that it depends on the object's size and
mass only through a combination that is proportional to its volume
density, $\rho$. That is, once $\rho$ is specified,
$\tff$ is independent of the object's mass (or size), implying that,
in a collapsing uniform-density sphere, all spherical shells reach the
center at the same time. This is equivalent to the well-known property
that, for spherically-symmetric perturbations of a uniform medium, the
growth rate increases with increasing wavelength, and thus the fastest
mode of collapse is an overall contraction of the medium
\citep{Tohline80, Larson85}.

However, this independence of $\tff$ from the actual physical
dimensions of an object of fixed volume density is only valid when the
object's extension $R$ is comparable in all three spatial dimensions (a ``3D
object''), because only in this case is the volume density of the object
given by eq.\ (\ref{eq:ro3D}).
%That is, for a spherical object with
%uniform density, the mass increases rapidly enough with radius that the
%acceleration it imparts to a spherical shell exterior to it causes the
%latter to reach the center at the same time as the internal mass (cf.\
%sec.\ \ref{sec:tff_3D}).
Instead, for nearly sheet-like (``2D'') or filamentary (``1D'') shapes,
the volume density $\rho$ is {\it not} proportional $M/L^3$, where $L$
here generically denotes the object's largest dimension. For these
morphologies, $\rho \propto M/(\ell L^2)$ or $\rho M/(L \ell^2)$,
respectively, where $\ell$ denotes the fixed, small dimension(s) of the
object. This is relevant for interstellar structures, since they are
often observed to have sheet-like or filamentary, rather than spherical,
morphologies \citep[e.g.][]{Bally+89, deGeus+90, HT03, Myers09,
Molinari+10, Andre+10}. This suggests that the free-fall timescale may
actually depend on the size of a non-spherical object in addition to
depending its volume density.

The gravitational stability of non-spherical structures has been
considered in earlier works \citep[e.g.,][]{Ledoux51, Larson85, Curry00},
but mostly considering {\it infinite} media and without discussing
collapse timescales. Finite-size non-spherical structures, and their
corresponding collapse times have only recently begun to be
considered. In particular, \citet{BH04} computed an approximation to the
free-fall time $\tffsh$ for finite-sized, infinitely thin circular
sheets of radius $R$, given by
\begin{equation}
\tffsh \approx \tffbh \equiv \sqrt{\frac{R}{\pi G \Sigma}},
\label{eq:tff_BH}
\end{equation}
where $\Sigma$ is the surface density (with units of mass per unit area)
of the sheet. 
%
%This approximation was obtained through a series
%expansion for the expression for the radial acceleration of a point near
%the periphery of the object, retaining only the linear term, and
%integrating to obtain the total collapse time.
%% It is noteworthy that this expression contains
%%the {\it surface} density $\Sigma$ and the radius of the sheet. 
%However, if we assume that the sheet has a small but finite thickness $h
%\ll R$ and uniform volume density $\rho$, so that $\Sigma = \rho h$, we
%can rewrite $\tffbh$ as
%%
%\begin{equation}
%\tffbh = \sqrt{\frac{A}{\pi G \rho}},
%\label{eq:tff_BH2}
%\end{equation}
%%
%where $A \equiv R/h$ is the {\it aspect ratio} of the sheet. 
It is noteworthy that, in this case,
the free-fall time exhibits a dependence on its size
in addition to depending on the (column) density.
%  \citet{BH04}
%also computed the acceleration for an end point of a thin filament,
%although they did not compute the corresponding free-fall time
%analytically.
Indeed,
numerical simulations of the collapse of large sheet-like clouds
containing many Jeans masses by
\citet{VS+07} exhibited collapse timescales significantly larger
than their corresponding three-dimensional free-fall time, as given by
eq.\ (\ref{eq:tff}). 
%Specifically, for gas at number density $n \sim 100
%\pcc$ and temperature $T \sim 100$ K,
%$\tffsph \sim 3$ Myr, while the clouds in the simulations collapsed
%on timescales of $\sim 15$ Myr.
Motivated by these realizations, in this paper we compute in detail the
free-fall time for sheet-like and filamentary structures. 
Note that \citet{Pon+11} have recently investigated
the free-fall timescales for sheet-like and filamentary geometries,
although they have focused on whether small-scale perturbations within
such structures have sufficient time to collapse before the whole
structure does so. Here we concentrate on a different question: 
whether
the collapse timescale for these geometries is longer, and by what
amount, compared to their spherical counterparts. Thus, our study and
that by \citet{Pon+11} can be considered as being complementary to each
other.

\section{The spherical case} \label{sec:tff_3D}

The standard calculation of the free-fall time for a uniform-density
sphere proceeds as follows \citep[e.g.,][]{BT87}. Consider a
uniform sphere of radius $R$, mass $M$, and volume density $\rho =
\rhosph(M,R) \equiv 3M/ 
4 \pi R^3$, which at time $t=0$ starts to contract under the action of
its self-gravity exclusively. The subindex ``3D'' denotes the
{\it function} that is used to calculate a quantity for a spherical (or
``3D'') geometry, which we distinguish from the {\it physical value} of the
quantity itself, denoted without subscript. At a certain later
time $t>0$, when the sphere 
has radius $r < R$, the velocity of a point at its periphery is
\begin{equation}
\frac{dr}{dt} = - \sqrt{2 G M \left(\frac{1}{r} - \frac{1}{R}\right)}.
\label{eq:vff3D}
\end{equation}
Introducing the non-dimensional variable $x\equiv r/R$, this is
equivalent to
\begin{equation}
\frac{dx}{dt} = - \sqrt{\frac{2 G M}{R^3} \left(\frac{1-x}{x}\right)}.
\label{eq:vff3D_adim}
\end{equation}
It is then clear that, because for a spherically symmetric uniform
object the mass $M$ increases as $R^3$, expression (\ref{eq:vff3D_adim})
depends on the mass and size of the sphere only through the volume
density they imply, $\rho = \rhosph$. Equation (\ref{eq:vff3D_adim}) can then
be integrated to yield
\begin{equation}
\tff =\tffsph(M,R) = \left(\frac{R^3}{2GM}\right)^{1/2} \int_0^1
\left(\frac{x}{1-x} \right)^{1/2} dx = \sqrt{\frac{3 \pi}{32 G \rho}},
\label{eq:tff3D_deriv}
\end{equation}
evidencing the independence of $\tff$ on the cloud's size or mass for a given
volume density.

\section{Circular sheet-like cloud} \label{sec:tff_2D}

Let us now consider the case of an infinitely thin circular sheet of
mass $M$, initial radius $R$, and uniform surface density $\Sigma=M/ \pi
R^{2}$. Here we follow closely the analysis by \citet[][hereafter
BH04]{BH04}. They calculated the radial acceleration experienced by the
periphery of the sheet when it has shrunk to radius $r<R$, under the
assumption that the surface density $\Sigma$ remains constant. This
assumption was made because they found that the maximum acceleration
occurs at the periphery of the sheet, and thus a dense contracting ring
forms in the periphery, while the inner surface density remains
essentially unchanged. This behavior has been observed in numerical
simulations by BH04 themselves and by \citet{VS+07}. With this
assumption, BH04 found
\beq
a_r = 4 G \Sigma \frac{R}{r}\left[K\left(\frac{r}{R}\right) - E\left(
\frac{r} {R} \right)\right],
\label{eq:a_r}
\eeq
where $K$ and $E$ are respectively the first and second elliptic
integrals. Upon a series expansion, eq.\ (\ref{eq:a_r}) becomes
\begin{equation}
\label{eq:gravity_sheet}
a_{r} =  \pi G \Sigma \left[ \frac{r}{R} +
\frac{3}{8}\left(\frac{r}{R}\right)^{3} +
\frac{45}{192}\left(\frac{r}{R}\right)^{5} + \dotsb \right]. 
\end{equation}
Note that the acceleration diverges at $r=R$, but BH04 noted that this
problem is eliminated when a finite sheet thickness is considered.

BH04 retained only the linear term in eq.\ (\ref{eq:gravity_sheet})
and computed the free-fall time by noting that $a_r = dv/dt =
1/2~dv^2/v~dt = 1/2~dv^2/dr$, where $v$ is the instantaneous radial
velocity. They thus found the instantaneous velocity as a function of
the instantaneous radius of the sheet, given by
\beq
v^2(r) = - \frac{2 \pi G \Sigma} {R} \int_R^r r^\prime dr^\prime =
\frac{\pi G \Sigma} {R} \left(R^2 - r^2\right),
\label{eq:vel}
\eeq
implying that the radial velocity at the end of the collapse is
\beq
v(r=0) = \sqrt{\pi G \Sigma R}.
\label{eq:v(0)}
\eeq

Taking this as a representative velocity for the entire collapse, BH04
then found a lower limit to the time for the sheet to shrink from $r=R$
to $r=0$, which we label $\tffbh$, given by
\begin{equation}
\label{eq:t1_BH}
\tffbh = \frac{R}{v(r=0)} = \sqrt{\frac{R}{\pi G \Sigma}},
\end{equation}
as anticipated in eq.\ (\ref{eq:tff_BH}). 

At this point we can extend the calculations by BH04 in two ways. First,
we consider clouds with small but finite thickness, so that it is
possible to define a {\it volume} density within them. We refer to this
as a ``quasi-2D'' geometry. Specifically, we rewrite the surface density
assuming a small but finite thickness $h$ such that $\Sigma=\rho h$,
where $\rho$ is the volume density. However, in this case the
volume density is not given by eq.\ (\ref{eq:ro3D}), but rather by $\rho
= \rhosh(M,R) = M/\pi R^2 h$. Here, the subindex ``2D'' now denotes the
relevant function to compute the quantity for a quasi-2D structure. 

Second,
instead of taking the final velocity as representative of the entire
collapse, we can obtain a more accurate expression by integrating eq.\
(\ref{eq:gravity_sheet}) from $R$ to $r$ to write an expresion for
$v(r)$, obtaining
\begin{equation}
\label{eq:vel_1}
v_{1}(r) = \sqrt{\frac{\pi G \rho}{A}\left( R^{2} - r^{2} \right)},
\end{equation}
where $A\equiv R/h$ is the {\it aspect ratio}. After a second
integration we obtain

\begin{equation}
\label{eq:radio_1}
r_{1}(t) = R~\sin{\left( \frac{\pi}{2} - \sqrt{\frac{\pi G
\rho}{A}}t \right)}, 
\end{equation}
where the subindex `1' denotes the assumption that the density
of the sheet internal to its periphery remains constant. Setting $r_1 =
0$, we obtain the corresponding free fall time as
\begin{equation}
\label{eq:time_1}
\tffone = \sqrt{\frac{A \pi}{4 G \rho}} = \sqrt{\frac{8A}{3}} \tffsph,
\end{equation}
where the second equality compares with the free-fall time
that would be obtained for a spherical structure with the same volume
density, explicitly exhibiting the extra factor $\propto \sqrt{A}$.

Let us now assume that, instead of remaining constant, the sheet's
density increases during the contraction so as to maintain the sheet's
mass constant; that is, $\Sigma \left(r(t)\right) = M/ \pi r(t)^{2} =
\rho\left(r(t) \right)h$, with $M=$ cst. Note that, in reality, the
sheet's mass does remain constant 
during the collapse, but it is not distributed uniformly on the
sheet. Instead, the mass external to $r$ is piled up at the
periphery. So, a point at the periphery sees this mass at the farthest
possible distance within the sheet, rather than seeing it uniformily
distributed over the sheet. Thus, the constant-mass assumption
overestimates the gravitational pull of the sheet on this point. On the
other hand, the constant-density assumption neglects the mass at the
periphery altogether. Thus, the two assumptions should bracket the real
situation (still within the linear approximation).

\begin{figure*}[p]
\begin{center}
  \includegraphics[width=0.5\linewidth]{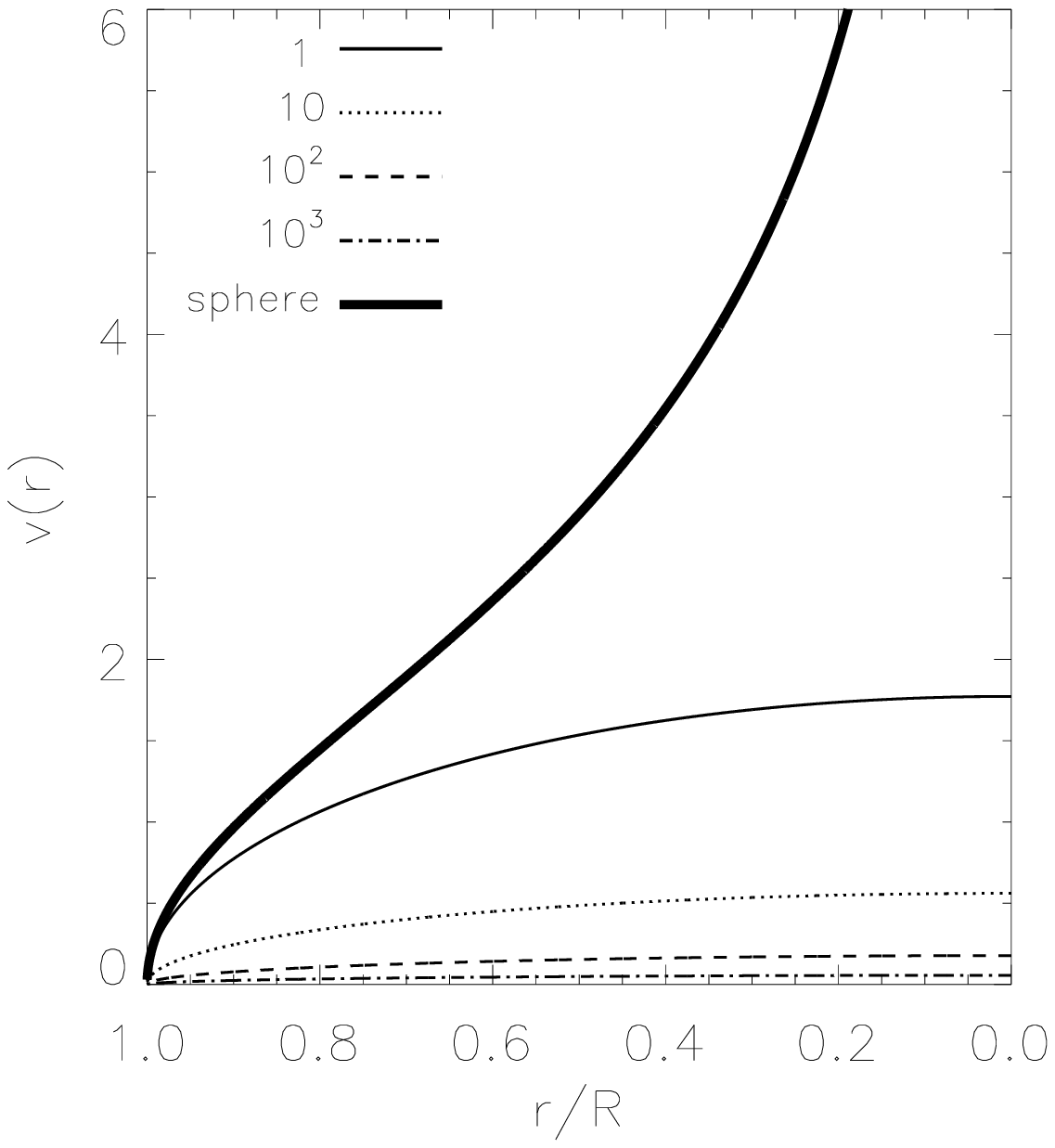}~
  \includegraphics[width=0.5\linewidth]{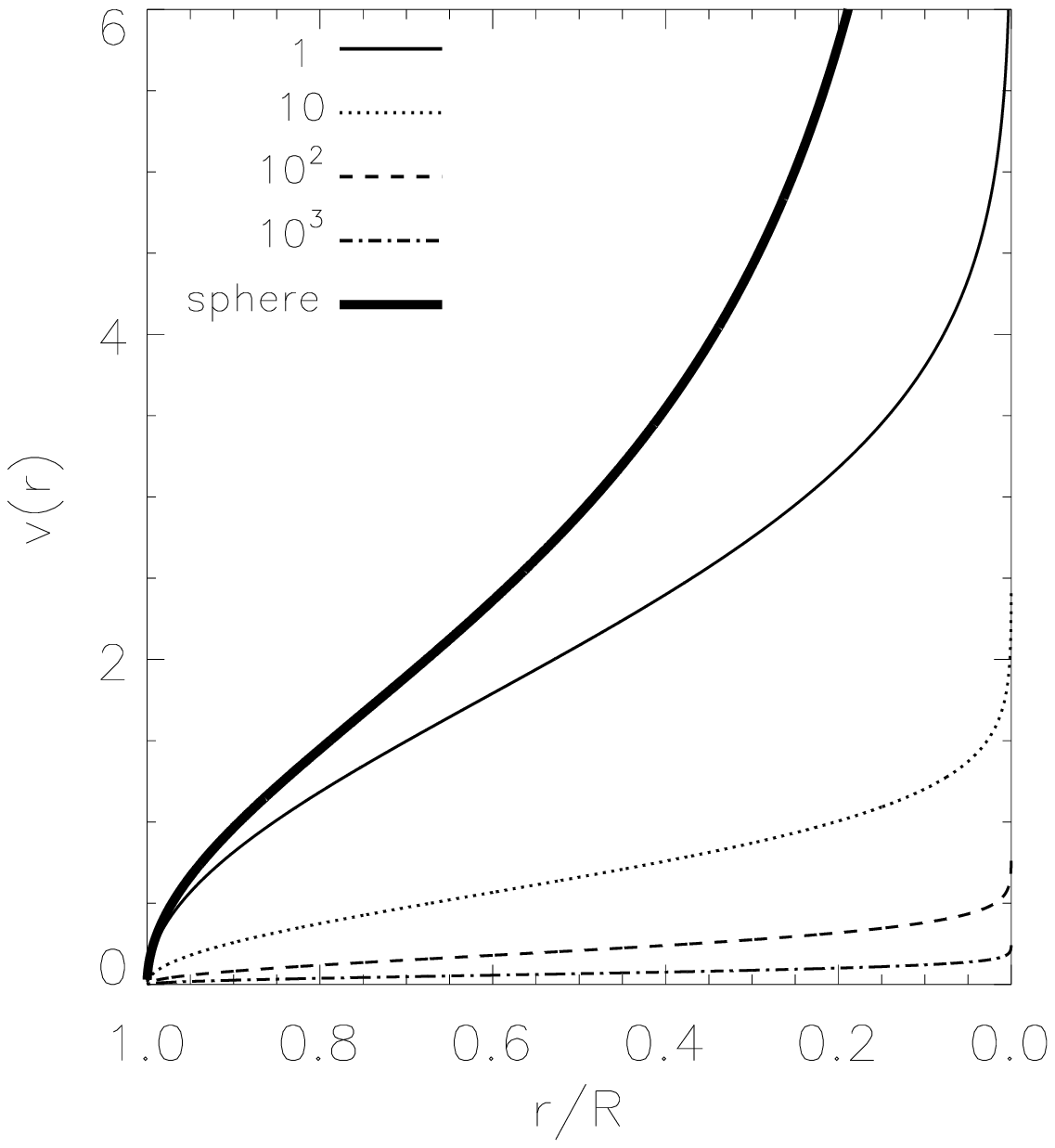}\\
  \includegraphics[width=0.5\linewidth]{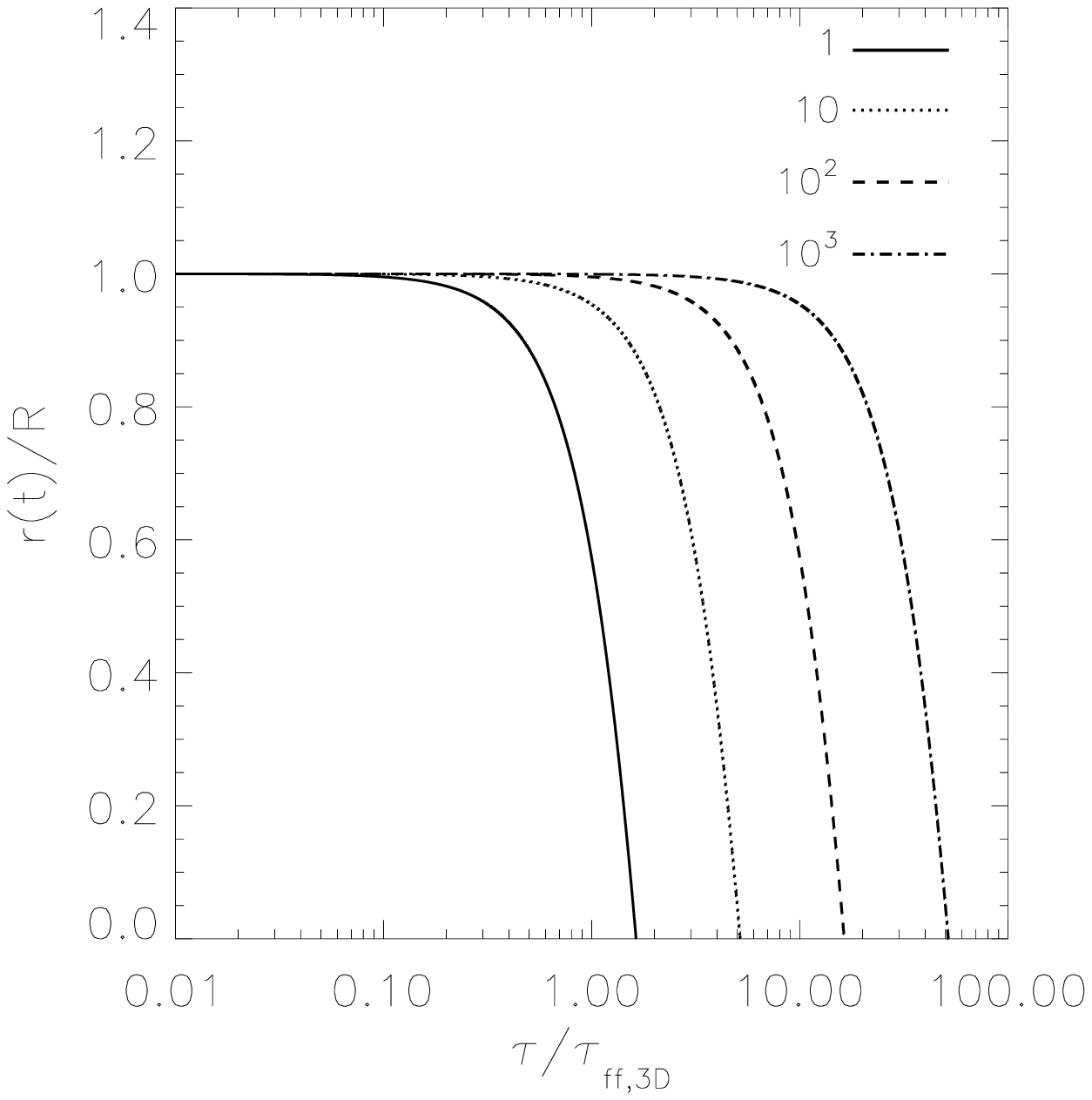}~
  \includegraphics[width=0.5\linewidth]{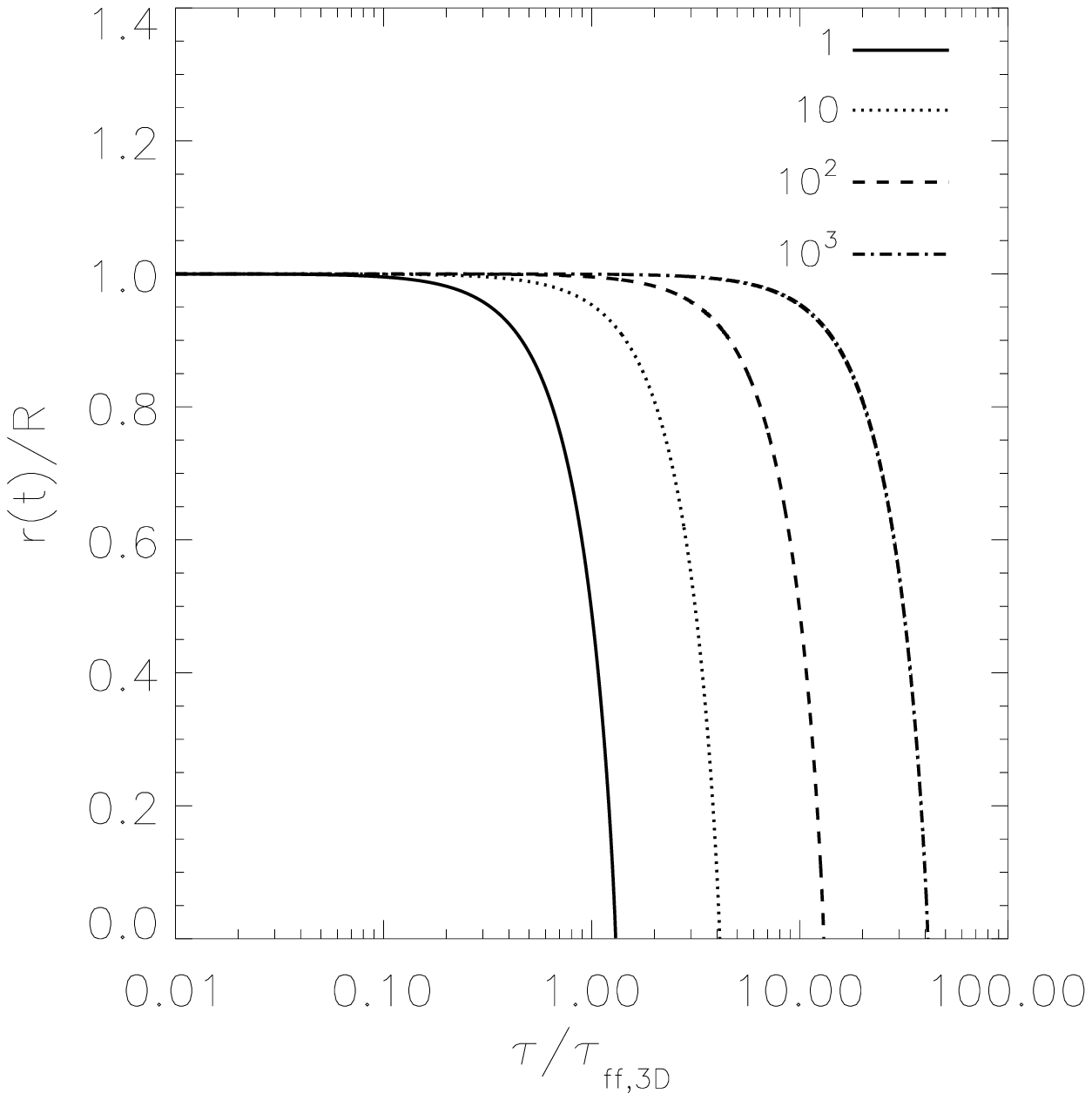} \caption{\textit{Top:}
  Evolution of the radial velocity of a sheet's periphery (shown as a
  function of its instantaneous radius, normalized to its initial value
  $R$) as it collapses, compared to the corresponding velocity for a
  spherical cloud (thick line). Note that the radius decreases to the
  right. \textit{Bottom:} Normalized position of the sheet's radius as a
  function of time, in units of the spherical free-fall
  time. \textit{Left:} Case 1. \textit{Right:} Case 2. The different
  lines show different values of $A$.}
\label{fig:vel_rad_sheet}
\end{center}
\end{figure*}

Under this assuption we can integrate eq.\ (\ref{eq:gravity_sheet})
from $R$ to $r$ to obtain a velocity function as
\begin{equation}
\label{eq:vel_den}
v_{2}(r) = \sqrt{2 \pi G \rho h R\ln{\left(\frac{R}{r}\right)}} = 
\sqrt{\frac{2 \pi G \rho R^{2}}{A} \ln{\left( \frac{R}{r} \right)}},
\end{equation}
where the subindex `2' denotes the case of a constant-mass assumption.
Upon a second integration, we obtain
\begin{equation}
\label{eq:radio_den}
r_{2}(t) = \frac{R}{\exp{\left[\mathrm{erf^{-1}}\left(   \sqrt{ \frac{2
G \rho}{A}}t  \right) \right]^2} }, 
\end{equation}
where $\mathrm{erf^{-1}}$ is the inverse error function. This gives a
free-fall time of
\begin{equation}
\label{eq:time_ff}
\tfftwo =  \sqrt{\frac{A}{2 G \rho}} = \sqrt{\frac{16 A}{3 \pi}}\tffsph,
\end{equation}
which is $\sim 25$\% shorter than $\tffone$ because the gravitational
acceleration is always larger, as the surface density increases
monotonically as the sheet contracts. The estimates for the free-fall
time in the two cases differ by factors of order unity at most. More
importantly, both of them have the same dependence on the aspect ratio
as $\sim\sqrt{A}$.  This dependence is fundamentally different from that
of the spherical (``3D'') case, given by eq.\ (\ref{eq:tff3D_deriv}), as
discussed in Sec.\ \ref{sec:intro}. This is illustrated in
Fig.\ref{fig:vel_rad_sheet}\,(\textit{Top panels}), which shows the
infall velocities as a function of the instantaneous radius,
corresponding to eqs.\ (\ref{eq:vel_1}) and (\ref{eq:vel_den}), taking
$R =\rho = G = 1$, for a range of values of $A$. Also shown is the infall
velocity for the spherical case. We see that the latter is always larger
than either $A$ values in both cases. This result is clearly a
consequence of the much smaller mass contained in a flat, essentially
``2D'', configuration than in a spherical (3D) one of the same volume
density. 

The {\it bottom panels} of Fig.\ \ref{fig:vel_rad_sheet} show the
sheet's radius as a function of time for the two density cases. As
expected, case 1, in which the cloud's mass is not conserved, but
rather decreases in time, undergoes a slightly slower collapse than
case 2. Nevertheless, we see that the difference in the collapse time
between the two cases is around $\sim25\%$, and so both are
qualitatively similar.

%Finally in Fig.\ \ref{fig:time1} we show the dependence of the
%free-fall time with the aspect ratio $A$ for both cases.

%
%\begin{figure*}[p]
%\begin{center}
%\includegraphics[width=0.48\linewidth]{fig2.ps}
%\caption{Ratio of each of the 2D free-fall times $\tffone$ and $\tfftwo$
%  to the 3D free-fall time, $\tffsph$, as a function of the aspect
%  ratio $A$.}
%\label{fig:time1}
%\end{center}
%\end{figure*} 

\section{Filamentary cloud} \label{sec:tff_1D}

We now consider the case of a uniform, cylindrical cloud of 
total length $2L$, radius $\calR \ll L$, and volume density $\rho =
\rholin(M,L) = M/\pi 
\calR^2 L$, where now the subindex `1D' denotes the function appropriate
for calculating a physical quantity in the quasi-1D case. Note that here
the filament's radius $\calR$ is the small, fixed dimension, and the
size variable is $L$.

Again, we start with the case in which the filament's
density remains constant. In this
case, the acceleration towards the filament's center at a
distance $l$ from the center is given by (BH04)
\begin{equation}
a(l) = \frac{1}{2} \frac{dv^2}{dl}   = -2 \pi G \rho \left[ 2l + \calR -
\sqrt{\calR^{2} + 4l^{2}}  \right].
\end{equation} 
Integrating this equation from $L$ to $l$, we obtain the radial
velocity after the filament has contracted from $L$ to $l$:
\begin{equation}
\label{eq:vel_cyl}
v(l) = \sqrt{4 \pi G \rho} \left[(L-l)(L+l+\calR) - \frac{L}{2}\sqrt{\calR^{2}
    + 4L^{2}} + \frac{l}{2}\sqrt{\calR^{2} + 4l^{2}} - \frac{\calR^{2}}{4}
  \ln{\left|\frac{\sqrt{\calR^{2} + 4L^{2}} + 2L}{\sqrt{\calR^{2} + 4l^{2}} +
      2l} \right|} \right]^{1/2}.
\end{equation}
If we now define the nondimensional parameters $A=L/\calR$ and $x=l/L$,
we can rewrite eq.\ (\ref{eq:vel_cyl}) as
\begin{equation}
\label{eq:vel_cyl1}
\frac{dx}{dt}= \frac{\pi}{2 \tffsph}\sqrt{\frac{3}{2}}
\left[(1-x)(1+ x + \frac{1}{A}) - 
  \frac{1}{2}\sqrt{\frac{1}{A^{2}} + 4} +
\frac{x}{2}\sqrt{\frac{1}{A^{2}} + 4x^{2}} - 
  \frac{1}{4 A^{2}} \ln{\left|\frac{\sqrt{\frac{1}{A^{2}} + 4} +
2}{\sqrt{\frac{1}{A^{2}} +  4x^{2}} + 2x} \right|} \right]^{1/2}.
\end{equation}
This equation again shows that the velocity depends on the
``aspect ratio'' $A = L/\calR$, besides the standard dependence on
density given by $\tffsph$. 

Let us now drop the constant-density assumption and assume instead
that the volume density increases as the filament contracts as $\rho(l)
= \rholin(M,l) = M /\pi \calR^{2}l$. In this case, we can perform the
same calculation for the velocity evolution, to obtain
\begin{equation}
\label{eq:vel_cyl2}
\frac{dx}{dt}= \frac{\pi}{2 \tffsph}\sqrt{\frac{3}{2}} \left[
  2(1-x) - \sqrt{\frac{1}{A^{2}} + 4} + \sqrt{\frac{1}{A^{2}} + 4
    x^{2}} - \frac{1}{A} \ln{\left| \frac{\sqrt{\frac{1}{A^{2}} + 4
        x^{2}} + \frac{1}{A}}{\sqrt{\frac{1}{A^{2}} + 4 } +
      \frac{1}{A}} \right|} \right]^{1/2}.
\end{equation}

\begin{figure*}[p]
\begin{center}
  \includegraphics[width=0.5\linewidth]{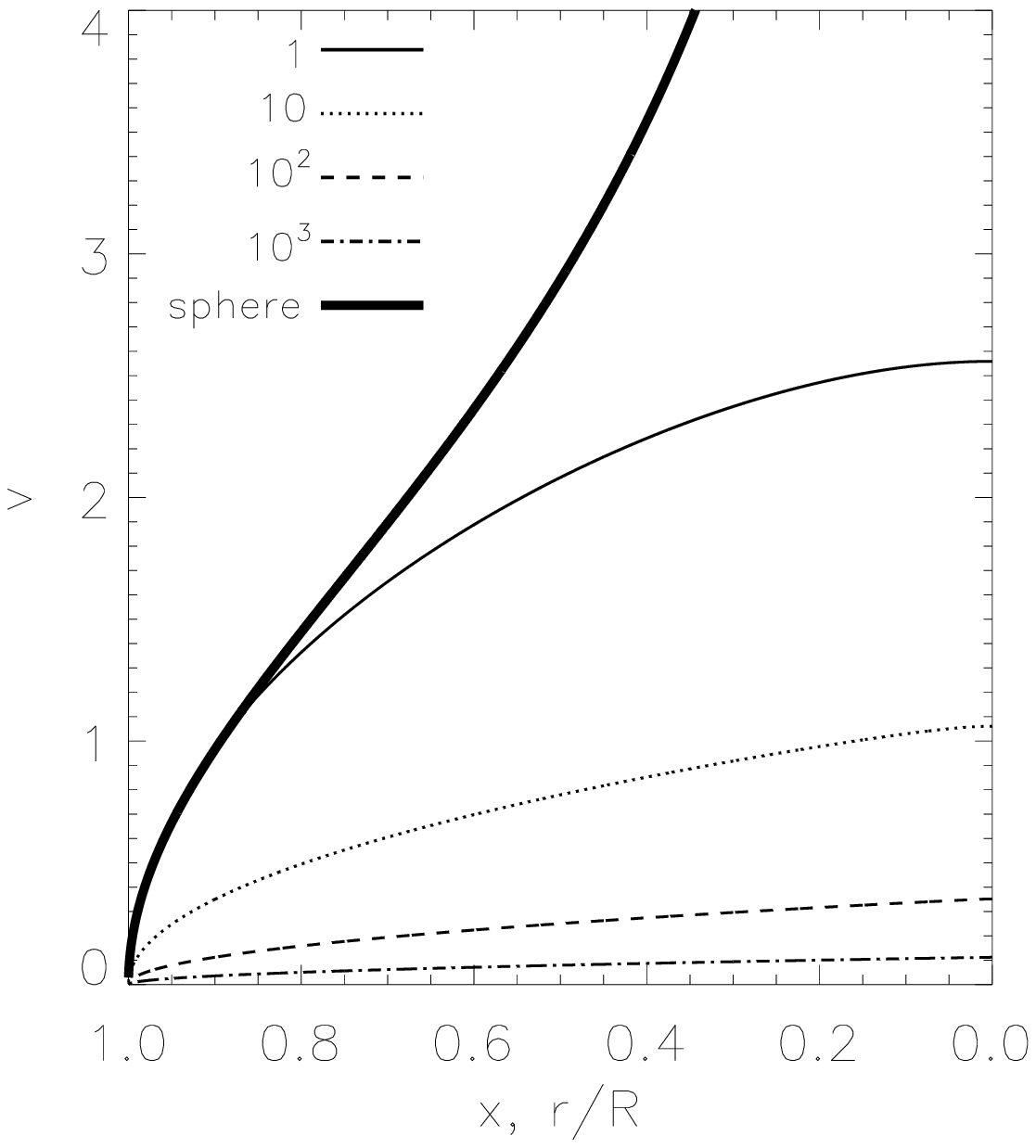}~
  \includegraphics[width=0.5\linewidth]{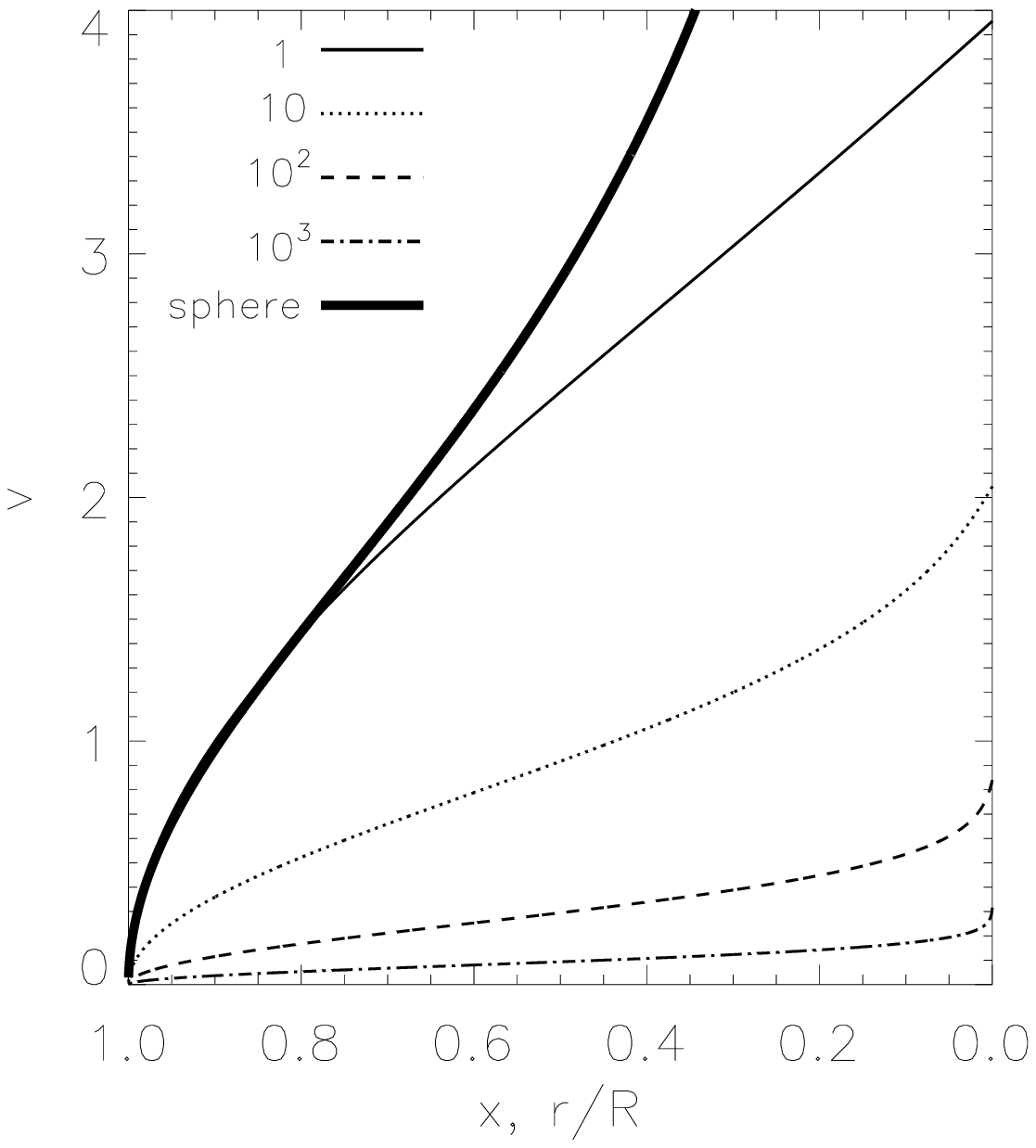}\\
  \includegraphics[width=0.5\linewidth]{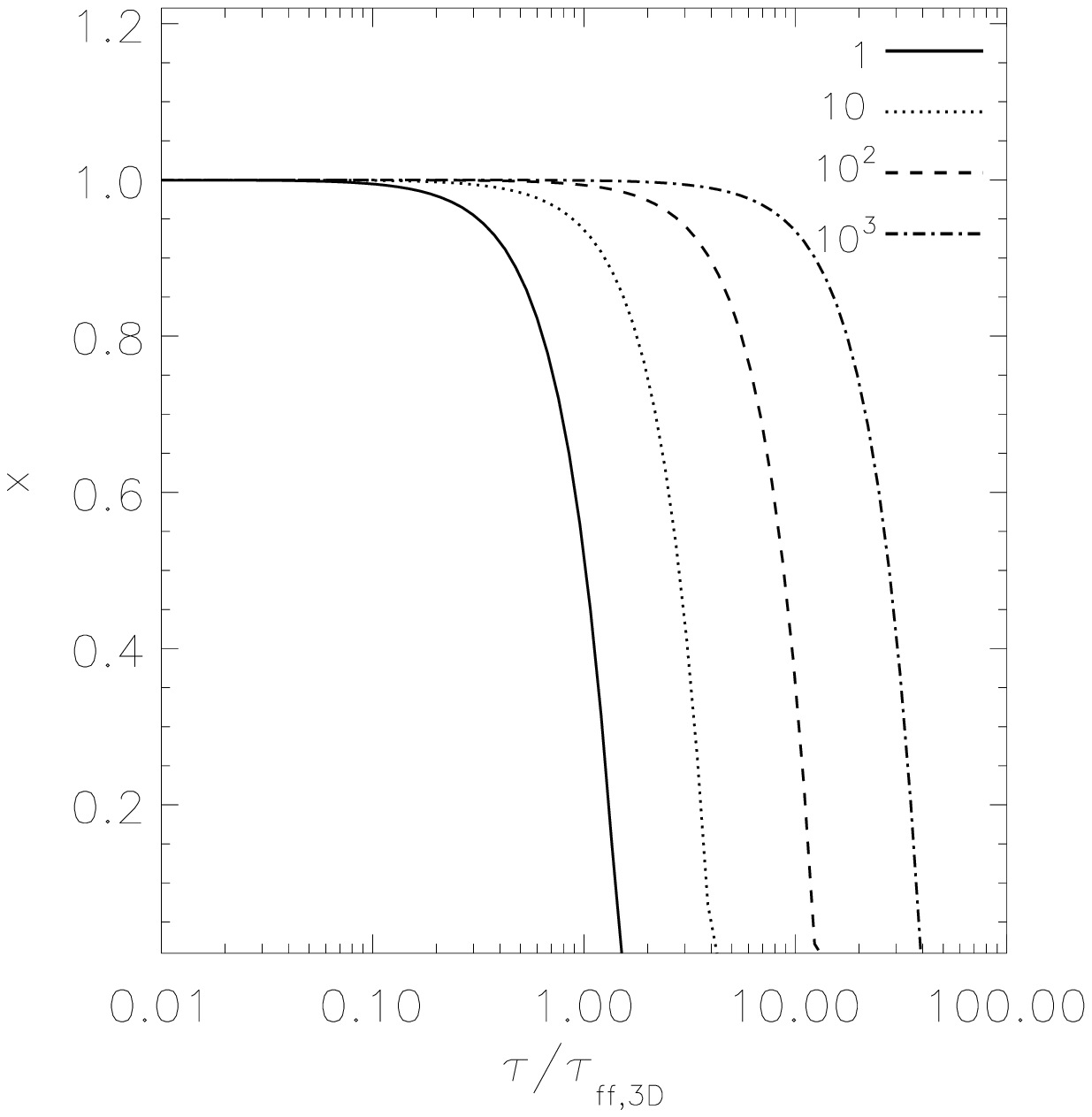}~
  \includegraphics[width=0.5\linewidth]{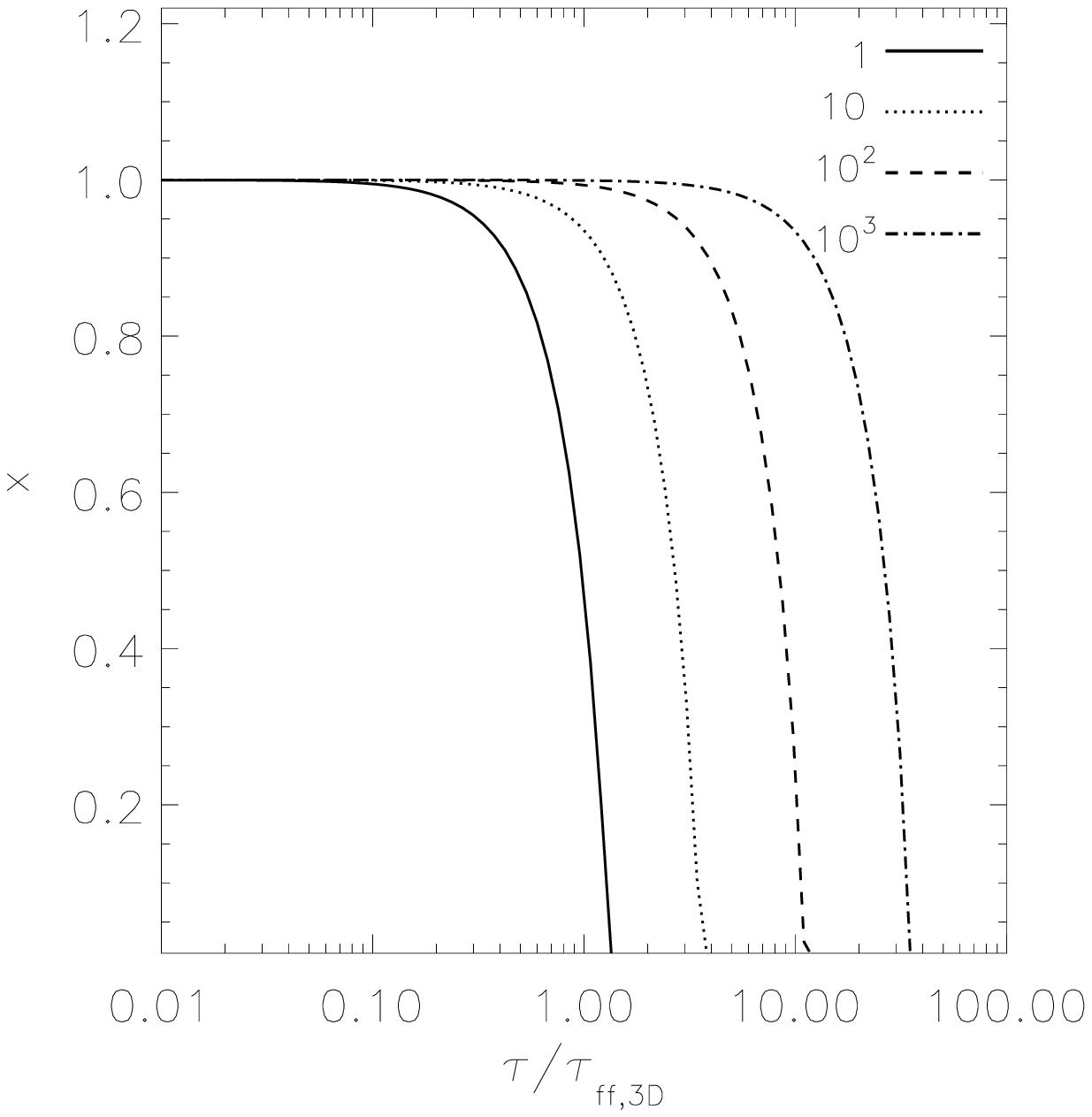}
\caption{\textit{Top:} Evolution of the edge velocity for a filamentary cloud,
  compared to the corresponding velocity for a spherical cloud (thick
  line). \textit{Bottom:} Normalized position of the filament's edge as
  a function of time, normalized to the spherical free-fall time
  estimate. \textit{Left:} Case 1, with $\rho =$ cst. \textit{Right:}
  Case 2, taking $\rho(l) = M/ \pi \calR^{2} l $. The different lines
  show different values of $A$.}
\label{fig:cil}
\end{center}
\end{figure*}

In Fig.\ \ref{fig:cil}\,(\textit{top panels}) we show the contraction
velocity as a function of $l$ for a range of values of $A$, for both
the constant-density and constant-mass cases.
In both, the final velocity reaches
smaller values as larger values of $A$ are considered.
The {\it lower panels} in Fig.\ \ref{fig:cil} show the position of the
filament's edge as a function of time, obtained from numerical
integration of eqs.\ (\ref{eq:vel_cyl1}) and (\ref{eq:vel_cyl2}). The
results are normalized to the spherical free-fall time. 

\begin{figure*}[p]
\begin{center}
\includegraphics[width=0.48\linewidth]{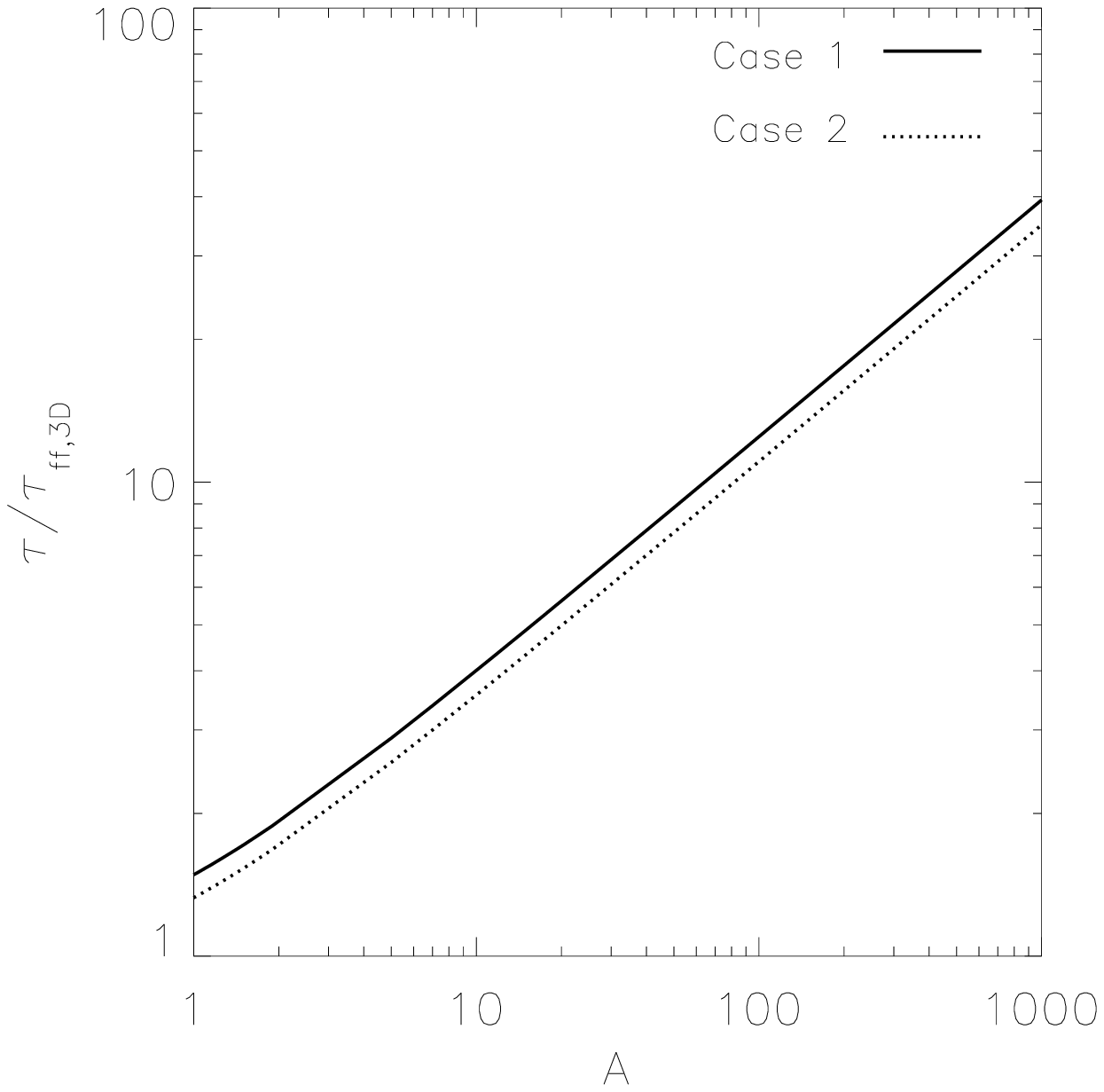}
\caption{Dependence of the free-fall time for a filamentary cloud on the
parameter $A$. The asymptotic behavior $\propto \sqrt{A}$ is seen to be
reached quickly as $A$ is increased.}
\label{fig:time2}
\end{center}
\end{figure*}

Expressions (\ref{eq:vel_cyl1}) and (\ref{eq:vel_cyl2}) are quite
complicated, and thus it is worthwhile to consider the limiting case of
large $A$, for which an analytical solution can be found. We obtain
\begin{equation}
\label{eq:time_cyl_ff_1A}
\tffone = \frac{2}{\pi} \sqrt{\frac{8A}{3}} \tau_{\mathrm{ff,3D}} 
\end{equation}
and
\begin{equation}
\label{eq:time_cyl_ff_2A}
\tfftwo = \sqrt{\frac{8A}{3\pi}} \tau_{\mathrm{ff,3D}},
\end{equation}
from which we see that, at large $A$, the free-fall time for
filaments also exhibits an additional factor $\propto \sqrt{A}$
(Fig. \ref{fig:time2}), similarly to the case for sheets.

\section{Discussion and applications} \label{sec:discussion}

\subsection{Implications for filamentary structure in molecular clouds}
\label{sec:impl_fil}

Our results have a number of implications for various aspects of
interstellar structure and star formation. First, the fact that both the
sheet-like and filamentary geometries have free-fall timescales larger
than those of their three-dimensional (spherical) counterpart with the
same volume density implies that any sheet or filament containing many
Jeans masses will collapse {\it later} than any approximately-3D clump
within it containing {\it one} Jeans mass. That is, the 3D clump will
collapse on roughly one spherical free-fall time, while the sheet or
filament around it will terminate its own collapse after a time that is
longer by a factor $\sqrt{A}$. This naturally explains the commonly
observed morphology of clumps immersed within, and accreting from,
filaments \citep[e.g.][]{Myers09, Purcell+09, Schneider+10, Pillai+11},
because the clumps evolve on a shorter timescale than their parent
filaments.

Specifically, our results can be applied to the filaments
observed in the Aquila rift and Polaris Flare regions by
\citet{Andre+10}. These authors report a typical filament width of
$2\calR \sim 10^4$ AU, or $\sim 0.05$ pc. In turn, filaments extending
for over 3 pc can be readily seen in their Fig.\ 1, implying aspect
ratios of up to $\sim 60$. From our Fig.\ \ref{fig:time2}, we see that
this in turn implies filament collapse times up to $\sim 10$ times
larger than the corresponding spherical one.  Moreover, \citet{Andre+10}
report a typical column density of the filaments of $N \sim 10^{22}
\psc$. We can thus estimate the mean volume density in the filaments as
$n \approx N/2
\calR \approx 6.5 \times 10^4 \pcc$, for which, assuming a mean
molecular mass $\mu = 2.36$, the {\it spherical} free fall time is $\tffsph
\approx 1.3 \times 10^5$ yr, and the Jeans length is $\LJ \approx 0.08$
pc. According to our results, a filament of length $2L = 3$ pc would thus
take a time $\sim 1.3$ Myr to collapse. But if it contracts towards its
center of mass, then quickly the central parts of the filament will
reach a volume density large enough that a segment of the filament of
length equal to its thickness will become Jeans unstable. Indeed, the
required volume density for the Jeans length to become $\LJ = 2\calR =
0.05$ pc is $n_{\rm J} \sim 1.8 \times 10^5 \pcc$, or only $\sim 2.8$ times
the mean volume density of the filament. At this point, the central
clump becomes locally Jeans unstable, and proceeds to collapse on a
timescale up to one order of magnitude shorter than the filament,
naturally producing a star-forming core with a filamentary ``appendix''
that accretes onto it. Moreover, this will cause the core to sustain
star formation for times significantly longer than its own (spherical)
free-fall time, since its gas supply is continuously replenished by the
accretion flow from the filament.

In turn, filaments may form from the collapse of sheets. As mentioned in
sec.\ \ref{sec:tff_2D}, during the collapse of a thin sheet, the
acceleration is maximal at the edge, and thus the collapse occurs from
the outside in, forming a dense ring in the periphery, which is
topologically equivalent to a filament. Such hierarchical fragmentation
from sheets to filaments has been suggested by several authors
\citep{SE79, GUW84, Hanawa+93, KP95}, and has also been readily
observed in numerical simulations \citep{Turner+95, BH04, HB07, VS+07,
Heitsch+09, Rosas+10, Prieto+11}. Thus, the collapse of giant molecular
clouds may proceed in a way that is anything but a monolithic,
three-dimensional collapse.

\subsection{Implications for the free-fall estimate of the SFR}
\label{sec:impl_SFcon}

A second implication of our results is that
they may contribute, at least partially, towards alleviating the well
known ``star formation conundrum''. The latter consists in that the
observed star formation rate (SFR), of a few $\msun$ yr$^{-1}$
\citep[e.g.,][]{Smith+78, Diehl+06, RW10},
is roughly two orders of magnitude lower than the simple estimate
obtained by dividing the total molecular mass in the Galactic ISM by the
{\it three-dimensional} free-fall time corresponding to the mean density
and temperature of the molecular gas \citep{ZP74}. This argument caused
the dismiss of the original proposal by \citet{GK74} that molecular
clouds should be in gravitational collapse, and was replaced by the
notion that the observed linewidths correspond to supersonic
microturbulence \citep{ZE74}. The latter was thought to provide support
against the clouds' self-gravity, a notion that persists until the
present. Later on, support by magnetic fields was considered as well,
and led to the notion that molecular clouds are in near virial
equilibrium, with SF proceeding at a much slower rate than free-fall,
only as allowed by mediation of ambipolar diffusion \citep[see, e.g.,
the reviews by][]{SAL87, Mousch91} and of local turbulent compression
\citep[e.g.,][]{VS+00, MK04}.

However, if molecular clouds and their substructure are indeed in near
free-fall, as suggested by various recent studies
\citep[e.g.,][]{BH04, VS+07, VS+10, HB07, PHA07, Galvan+09,
Schneider+10, Csengeri+11a, Csengeri+11b}, then the SF conundrum
reappears as a challenge to theoretical models of free-falling clouds
\citep[e.g.,][]{ZV11}. Our results may imply that, if the cold
(molecular and atomic) gas is distributed in sheets and filaments rather
than in three-dimensional structures, as suggested by various
observational studies \citep[e.g.,][]{Bally+89, deGeus+90, HT03,
Molinari+10, Andre+10}, then the 3D value of the free-fall time is an
underestimate to the real value, implying that the 3D free-fall SFR
value overestimates its actual value, and that the severity of the SF
conundrum may be reduced by up to one order of magnitude.

Of course, the correction provided by our results is not expected to
provide a {\it full} solution to the SFR conundrum, since it is unlikely
that sheets and filaments have the required aspect ratio values of $A
\sim$ a few $\times 10^3$ to completely account for a
two-order-of-magnitude discrepancy between the observed and the
free-fall estimate values of the SFR (see Figs.\
\ref{fig:vel_rad_sheet} and \ref{fig:cil}). Besides, the
presence of even weak magnetic fields (implying supercritical
mass-to-magnetic flux molecular clouds) and stellar feedback are
expected to reduce the SFR beyond whatever is achieved by the possible
low dimensionality of the clouds. Nevertheless, our result may provide a
reduction of the discrepancy factor between the observed SFR and its
free-fall estimate by factors ranging from a few to almost one order of
magnitude, since observed aspect ratios of sheets and filaments in the
ISM can reach values of up to $10^2$ \citep{HT03, Molinari+10}.

\section{Summary and Conclusions} \label{sec:concl}

%In this contribution, 
We have computed the free fall time for two
geometries that depart from spherical symmetry, namely, thin circular
sheets and long filaments. In both cases, we considered two different
behaviors for the volume density, one assuming that it remains constant
through the collapse and the other assuming that it is the
mass that remains constant, and the density increasing as the
object collapses. These two cases should bracket the actual
collapse timescale.

We have parameterized the problem by the aspect ratio $A$ of the object,
defined either as the ratio of the cloud's radius to its (small)
thickness ($A=R/h$) in the case of sheets, or as the ratio of the
cloud's half-length to its (small) radius ($A=L/\calR$), in the case of
filaments. Our calculations assume that the clouds contract only along
their largest dimension (radially for sheets, and longitudinally for
filaments), and thus our results are more accurate as larger values of
$A$ are considered. The value $A=1$ is a frontier value between the two
symmeties, although it is also the value for which our calculations are
in largest error, since three-dimensional contraction should ensue in
that case, with the free-fall time being given by $\tffsph$.
Nevertheless, even in this extreme case, our calculations only deviate
from the 3D value by factors of order unity.

For both cases, we found that the collapse time increases as
$A^{1/2}$. In particular, this implies that the collapse time of a thin
sheet is of the order of that of a sphere with much lower volume
density, so that it is its {\it column} density that is the same as that
of the sheet's.  This result has two important implications for the
structure of molecular clouds and star formation. First, it naturally
explains the common morphology observed in molecular clouds, where
star-forming or pre-stellar clumps are embedded within filaments that
appear to be accreting onto them. According to our results, this is a
natural consequence of the longer free-fall time for a filament than for
any clump-like structure within it that contains enough mass to be
itself collapsing. Because the filament has a longer collapse time, it
will continue to accrete onto the locally collapsing 3D clump after the
latter has managed to increase its density by a large enough amount as
to become distinguishable from the filament.

Second, it may imply that, if molecular gas in the Galaxy is distributed
in primarily low-dimensional structures such as sheets and filaments,
then the so-called SFR conundrum may not be as strong as it is normally
stated, because the relevant free-fall time for the cold gas in the
Galaxy may be longer than has been considered. However, we expect that
this effect is likely to only account for a fraction
of the discrepancy between the observed and the free-fall prediction for
the SFR, since the required aspect ratios to fully account for the
conundrum would be too large ($A \sim$ a few $\times 10^3$), and besides
other effects 
are known to contribute to reduce the SFR, such as magnetic fields and
stelllar feedback. In any case, our results suggest that determining the
topology of molecular clouds is important for estimating their true
expected collapse timescales.

\acknowledgements 
We acknowledge useful comments from an anonymous referee, which helped
improving the clarity and scope of the paper. This work has received
financial support from grants CONACYT 102488 to E.V.-S., UNAM-DGAPA
grant PAPIIT IN106511 to G.C.G, and a fellowship from CONACYT-SNI to
J.A.T.

\end{document}